# Stability of low-carrier-density topological-insulator $Bi_2Se_3$ thin films and effect of capping layers

Maryam Salehi[1], Matthew Brahlek[2, #], Nikesh Koirala[2], Jisoo Moon[2], Liang Wu[3], N. P. Armitage[3] and Seongshik Oh[2,*]

[1]Department of Materials Science and Engineering, Rutgers, The State University of New Jersey, Piscataway, New Jersey 08854, U.S.A.

[2]Department of Physics and Astronomy, Rutgers, The State University of New Jersey, Piscataway, New Jersey 08854, U.S.A.

[#]Current address: Department of Materials Science and Engineering, Pennsylvania State University, University Park, Pennsylvania 16802, USA

[3]The Institute for Quantum matter, Department of Physics and Astronomy, The Johns Hopkins University, Baltimore, Maryland 21218, U.S.A.

*Electronic mail: ohsean@physics.rutgers.edu

**Abstract:**

**Although over the past number of years there have been many advances in the materials aspects of topological insulators (TI), one of the ongoing challenges with these materials is the protection of them against aging. In particular, the recent development of low-carrier-density bulk-insulating $Bi_2Se_3$ thin films and their sensitivity to air demands reliable capping layers to stabilize their electronic properties. Here, we study the stability of the low-carrier-density $Bi_2Se_3$ thin films in air with and without various capping layers using DC and THz probes. Without any capping layers, the carrier density increases by ~150% over a week and by ~280% over 9 months. *In situ*-deposited Se and *ex situ*-deposited Poly(methyl methacrylate) (PMMA) suppresses the aging effect to ~27% and ~88% respectively over 9 months. The combination of effective capping layers and low-carrier-density TI films will open up new opportunities in topological insulators.**

Topological insulators (TIs) are new class of electronic materials which are insulating in the bulk, but metallic on the surfaces with linear Dirac-like surface states, which are robust against any defects and disorders as long as the time reversal symmetry is preserved[1-3]. $Bi_2Se_3$, an archetypal 3D topological insulator[4,5], has garnered a lot of attention over the past number of years. However, degradation and aging in air over time limit the realization of $Bi_2Se_3$-based devices. There have been transport and spectroscopic studies on how the properties of $Bi_2Se_3$ evolve when it is exposed to atmospheric gases[6-8]. In regular $Bi_2Se_3$ thin films grown on different substrates including sapphire[9], silicon[10], and amorphous silicon oxide[11], the surface Fermi level generally resides above the bottom of bulk conduction band and their sheet carrier densities are generally much higher than $\sim 1 \times 10^{13}$ cm$^{-2}$ with the bulk state of the samples metallic. Due to the initial high carrier densities, the carrier densities remain relatively stable in air with the change in carrier densities limited only to a few tens of percent[7,12]. The relative stability of these high-carrier-density samples has helped attaining numerous consistent results across different probing techniques even without any special care to protect the surface layers[9,13].

In contrast, recently developed low-carrier-density bulk-insulating $Bi_2Se_3$ films[14,15] exhibit quite significant aging effect in air. The as-grown sheet carrier densities of these samples are about an order of magnitude lower than the previous generation samples. This order of magnitude decrease in carrier density implies that the few tens of percent change in carrier density of the previous generation films in air would amount to a few hundreds of percent in these new films, if we assume that similar absolute level of aging effect occurs for both types of samples. Moreover, as we discussed earlier, in order to ensure the bulk insulating character of these low-carrier-density samples, it is critical to keep the sheet carrier densities substantially below $1 \times 10^{13}$ cm$^{-2}$ and maintain upward (instead of downward) band bending near the surfaces[16]. Below, we show that in order to stabilize the low-carrier-density $Bi_2Se_3$ films, it is essential to use proper capping layers. Among the three capping layers we studied, *in situ*-deposited $In_2Se_3$, *in situ*-deposited Se and *ex situ*-deposited PMMA, only Se and PMMA exhibited good stabilization effect.

The low-carrier-density Bi$_2$Se$_3$ films used for this study were grown on (Bi$_{1-x}$In$_x$)$_2$Se$_3$ buffer layers on sapphire substrate using a custom-designed molecular beam epitaxy (MBE) system by SVT Associates; details will be published elsewhere[15]. As for capping layers, 100 nm amorphous Se and 50 nm amorphous In$_2$Se$_3$ were deposited in situ at room temperature, whereas ~200 nm PMMA was spin-coated ex situ, followed by baking process at different temperatures. Both capped and uncapped films were measured for over 280 days. Standard Hall effect measurements were carried out at room temperature in air in magnetic field up to 0.6 T with the standard van der Pauw geometry, which uses indium wires pressed on the four corners of the square sample as contact leads. Sheet carrier density ($n_{sheet}$) is extracted from $n_{sheet} = (e\, dR_{Hall}/dB)^{-1}$ where $e$ is the electronic charge and $dR_{Hall}/dB$ is the slope of Hall resistance as a function of magnetic field $B$. Mobility is calculated using $\mu = (e\, R_{sheet}\, n_{sheet})^{-1}$ at zero magnetic field. In order to study the evolution of the transport properties over time, $n_{sheet}$ and $\mu$ were measured continuously for more than 36 hours immediately after the sample was taken out of the MBE chamber. After that, they were measured almost every day for 7 days, and then intermittently at couple-of-day intervals for the following ~9 months.

The continuous measurement of 30 QL (1QL ≈ 1 nm) uncapped sample for 40 hours (Fig. 1(a)) shows a telling decrease in $n_{sheet}$ (~30% drop from the initial value) and a simultaneous increase in $\mu$ at around hour 16, which is also observable in other uncapped samples. Similar short-time decrease in carrier density was observed in our previous study of high-carrier-density Bi$_2$Se$_3$ samples[7]. The trigger point of this decrease depends on ambient conditions, such as relative humidity and temperature. This shows that there are competing processes such that there is a short-time window in which the sample properties improve, and after that it gets worse rapidly until the carrier density reaches twice the initial value after a day. This drastic degradation necessitates an effective capping layer for stabilizing the properties of these thin films.

Figure 1(b) summarizes the change in carrier density for different capping layers in comparison with the uncapped sample for over 280 days, indicating different short-term and long-term behaviors in the films. Clearly, the uncapped sample shows a rapid degradation for ~80 days with 280% increase in $n_{sheet}$. However after that, film properties recover to some extent presumably because the charge exchange process between the film and air balances out. This recovery happens for all capping layers at some point, except that for the uncapped sample it is more noticeable since it recovers after a huge change in $n_{sheet}$. Another comparison in Fig. 1(c) shows the change in $\mu$ for all the capping layers along with the uncapped sample. It is clear that the mobility of the uncapped sample has a sharp drop in short time, and it gets slowly recovered after ~100 days. This is also true for capped samples, but the change in uncapped sample is more conspicuous.

$In_2Se_3$-capping layer makes the electrical properties of the film significantly worse immediately (Figs. 1(b) and 1(c)). The $In_2Se_3$-capped sample has initially almost two times higher $n_{sheet}$ compared to the uncapped sample. For depositing $In_2Se_3$ on $Bi_2Se_3$, selenium shutter was open for a few seconds prior to opening indium shutter to ensure Se rich environment in the vicinity of the substrate surface before indium atoms arrive; we tried various other schemes of depositing $In_2Se_3$ but all resulted in similarly degraded transport properties of $Bi_2Se_3$ films. One possible explanation for this drastic degradation is that indium tends to bond with selenium from the $Bi_2Se_3$ rather than the surrounding selenium, hence this mechanism leaves behind more selenium vacancies in the film. The result from $In_2Se_3$-capping layer shows that although superficially this capping layer may seem compatible with $Bi_2Se_3$ because of its similar structure, it does not function as an effective capping layer.

PMMA-capping, helps to suppress the drastic aging effect. Moreover, the coating and removing of this commonly used polymer is straight forward. Due to ex situ deposition, the transport results before and after PMMA coating could be compared. The films were exposed to air less than 10 minutes prior to PMMA-capping. Figure 2 summarizes $n_{sheet}$ and $\mu$ for different temperatures of PMMA baking. The initial values before PMMA-capping are also marked in the plots. There is a sudden increase in $n_{sheet}$ right after PMMA coating compared to the value before capping, which might be due to the delay in capping and/or curing of the PMMA layer. However sometime after coating (Fig. 2(a)), there is always a decreasing trend in $n_{sheet}$ until it reaches a minimum close to the original $n_{sheet}$. This is probably due to the fact that the solvent in the PMMA dries out over time. The inset graph in Fig. 2(a) shows the percentage change in $n_{sheet}$ right after PMMA-capping for each baking temperature compared to the original value before capping. This shows that for higher temperature baking, the amount of sudden change in the value of $n_{sheet}$ is lower, and this is probably due to the fact that harder baking reduces residual solvent and the density of pores in the PMMA-capping. Although compared to the uncapped sample, PMMA-capping makes the aging much slower, PMMA-capped sample goes through degradation before recovering again. This degradation is likely due to ambient species eventually reaching the film through PMMA layer, or due to change in PMMA properties over time. The start and peak of degradation process (Fig. 2(a)) depends on PMMA baking temperature, which presumably corresponds to the amount of leftover solvent in PMMA. For the case of change in $\mu$, the mechanism is more complicated but it is clear that the mobility of the sample measured right after PMMA coating is always lower than the value before capping, which might be due to delay in PMMA-capping or heat from the baking process. The mobility values eventually recover after a decreasing trend, which is consistent with our previous discussion.

Among the capping layers, selenium gives the best result in terms of stability of $n_{sheet}$ for both short and long-time compared to the uncapped sample. Our comprehensive study on Se-capped sample for more than 280 days using DC transport measurements demonstrates the remarkable effectiveness of this capping. It should be noted that the in situ deposition of this capping guarantees no direct exposure of the film's surface to air. Figure 1(a) shows that the change in $n_{sheet}$ is less than 27% over the course of 280 days, compared to uncapped sample, which confirms that the Se-capping efficiently protects the sample against environmental doping. Figure 1(d) shows the absolute change in daily $n_{sheet}$ value compared to the initial one for different Se-capping thickness for several days. For a better comparison, the values of the uncapped sample are plotted, too. The change in $n_{sheet}$ approximately follows an exponentially-saturating behavior: $C(1 - e^{-Dt})$, where $t$ is time. The long-time saturation value $C$ and the exponential rate $D$ are shown in the inset of Fig. 1(d) for both uncapped and Se-capped samples. The fact that $D$ is almost constant for all selenium thicknesses and is quite different for the uncapped sample, implies that *D* is a material constant, independent of the capping thickness. On the other hand, the value for *C* depends on the thickness of capping layer, which suggests that the thickness of capping layer determines the final level of contamination that reaches the underlying $Bi_2Se_3$ layer. Moreover, Se-capping was found to be very effective for surface sensitive techniques such as scanning tunneling microscopy (STM) or angle-resolved photoemission spectroscopy (ARPES); it has been shown by STM that a quick ion milling, prior to desorbing the Se-capping by heat, leaves behind an ultra-clean $Bi_2Se_3$ surface to work with[17].

Time-domain terahertz spectroscopy (TDTS) results for uncapped, PMMA- and Se-capped samples support the DC transport data. The TDTS measurements were performed at the Johns Hopkins University on samples grown at Rutgers University. The uncapped samples were measured a day (after one-day shipping) and a week after the growth day. A sharp change in the real conductance as a function of THz frequency (Fig. 3(a)) is noticeable in this short time. The real conductance for PMMA- and Se-capped samples was measured a day, a month and two months after the growth day (Figs. 3(b) and 3(c), respectively). The small change in the real conductance spectra for 20 nm Se-capping after 2 months aging confirms the effectiveness of this capping layer in stabilizing the sample's properties.

The carrier density of each topological surface state (TSS) can be calculated based on $n_{sheet} = k_F^2/4\pi$. An effective transport mass $m^* = \hbar k_F/v_F$ can still be defined for massless Dirac fermions where the Fermi velocity is determined by $v_F = \partial E_F/\hbar \partial k$. In our analysis, we will consider up to quadratic dispersion for surface states ($E_F = Ak_F + Bk_F^2$, where $A = 2.02$ eVÅ and $B = 10.44$ eVÅ$^2$) and model the top and bottom surface states as identical with the same carrier density as done previously[18,19]. Considering the TSS dispersion for surface states up to quadratic correction, the Drude spectral weight $\omega_{pD}^2 d$ (proportional to total areal under real conductance at zero field $G_{D1}$) can be expressed in terms of Fermi wave-vector $k_F$.

$$\frac{2}{\pi\epsilon_o}\int G_{D1} d\omega = \omega_{pD}^2 d = \frac{k_F(A+2Bk_F)e^2}{2\pi\hbar^2\epsilon_0} = \frac{n_{sheet}e^2}{m^*\epsilon_o} \qquad (1)$$

, where $\epsilon_o$ is the free-space permittivity, *e* is the electronic charge and *d* is the film thickness.

In Drude-Lorentz fit, the conductance spectra can be well fit by an oscillator model with a Drude term describing free electron-like motion, a Drude-Lorentz term modeling the phonon, and a lattice polarizability $\epsilon_\infty$ term that originates from absorptions outside the spectral range.

$$G(\omega) = \left[ -\frac{\omega_{pD}^2}{i\omega - \Gamma_D} - \frac{i\omega \omega_{pDL}^2}{\omega_{DL}^2 - \omega^2 - i\omega \Gamma_{DL}} - i(\epsilon_\infty - 1) \right] \epsilon_o d, \qquad (2)$$

, where $\omega_{pD}$ and $\Gamma_D$ are Drude plasma frequency and scattering rate. $\omega_{pDL}$, $\omega_{DL}$, and $\Gamma_{DL}$ are, respectively, the plasma and center frequencies, and scattering rate of optical phonons. DL stands for Drude-Lorentz.

The fit provides Drude plasma frequency $\omega_{pD}$ (which gives the spectral weight $\omega_{pD}^2 d$) and scattering rate $\Gamma_D$. By using the spectral weight and equation (1), we can first calculate $k_F$, then $n_{\text{sheet}}$ and effective mass $m^*$. Mobility is estimated afterwards by $\mu = e/(2\pi m^* \Gamma_D)$. Figures 3(d)-(g) show the change in spectral weight and scattering rate as well as the extracted $n_{\text{sheet}}$ and $\mu$ for fresh and aged samples, and confirm that Se-capping has the most stabilizing property. Here, it is important to note that even if we assumed a single Drude term with the surface state character to extract the transport parameters in Figs. 3(d)-(g), the effect of capping layers are obvious in the raw data (Figs. 3(a)-(c)) not only for the free-electron spectrum toward the low frequencies but also for the phonon peak. It is also worth noting that with higher thickness for Se-capping and higher baking temperature for PMMA-capping, the stability would have been even better.

To conclude, we have studied short- and long-time stability of low-carrier-density $Bi_2Se_3$ films with various capping layers. Both Se and PMMA capping layers help stabilizing the properties of $Bi_2Se_3$ films; they have additional advantage of easy removability for follow-up experiments. Another recently-found promising capping layer for $Bi_2Se_3$ is $MoO_3$[20]; it will be interesting to see how its long-term stability compares with Se and PMMA. Our study sheds light on the importance of capping layers on low-carrier-density TI films. Combination of effective capping layers and low-carrier-density TI films may open routes to previously unexplored regime of topological insulators.

**Acknowledgement**

This work was supported by ONR (N000141210456), NSF (DMR-1308142), and Gordon and Betty Moore Foundation's EPiQS Initiative (GBMF4418). The TDTS work (JHU) was additionally supported by Gordon and Betty Moore Foundation (GBMF2628).

**Figure captions**

**FIG. 1.** Aging effect on room temperature transport properties of 30 QL $Bi_2Se_3$ films with and without various capping layers. (a) $n_{sheet}$ and $\mu$ for an uncapped $Bi_2Se_3$ film over 40 hours. (b) $n_{sheet}$ and (c) $\mu$ vs. time for $Bi_2Se_3$ films with and without different capping layers over ~9 months. (d) Absolute change in $n_{sheet}$ vs. time for $Bi_2Se_3$ films with different thicknesses of Se-capping over several days. Dashed curves are the $C(1 - e^{-Dt})$ fittings. Inset shows the fitting parameter C and D for uncapped (open) and Se-capped (filled) films.

**FIG. 2.** Baking temperature dependence of aging effect for room temperature transport properties of 30 QL $Bi_2Se_3$ films with PMMA-capping. (a) $n_{sheet}$ and (b) $\mu$ vs. time for $Bi_2Se_3$ films with different baking temperatures of PMMA-capping. The open symbols show $n_{sheet}$ and $\mu$ values for as-grown samples before PMMA is coated and baked for each baking temperature, which are shifted to the left for clarity. The inset in (a) shows the percentage change in carrier densities right after PMMA-capping compared to before. The data points in (a) overlap due to similar carrier densities of the as-grown films.

**FIG. 3.** Aging of $Bi_2Se_3$ films measured by time domain THz spectroscopy at 5 K. Real conductance vs. THz frequency for 16 QL $Bi_2Se_3$ (a) with no capping, (b) with PMMA-capping (baked at 100 ºC), and (c) with 20 nm Se-capping, measured at different times after growth. The dashed lines are the Drude-Lorentz fits and the peak at ~1.9 THz is the bulk phonon mode of $Bi_2Se_3$. (d) Spectral weight, (e) Scattering rate, (f) $n_{sheet}$ and (g) $\mu$ vs. time for different capping layers: the latter two are derived from the former two quantities. Error bars represent the uncertainties of the parameters in the fits.

**FIG. 1 (2column)**

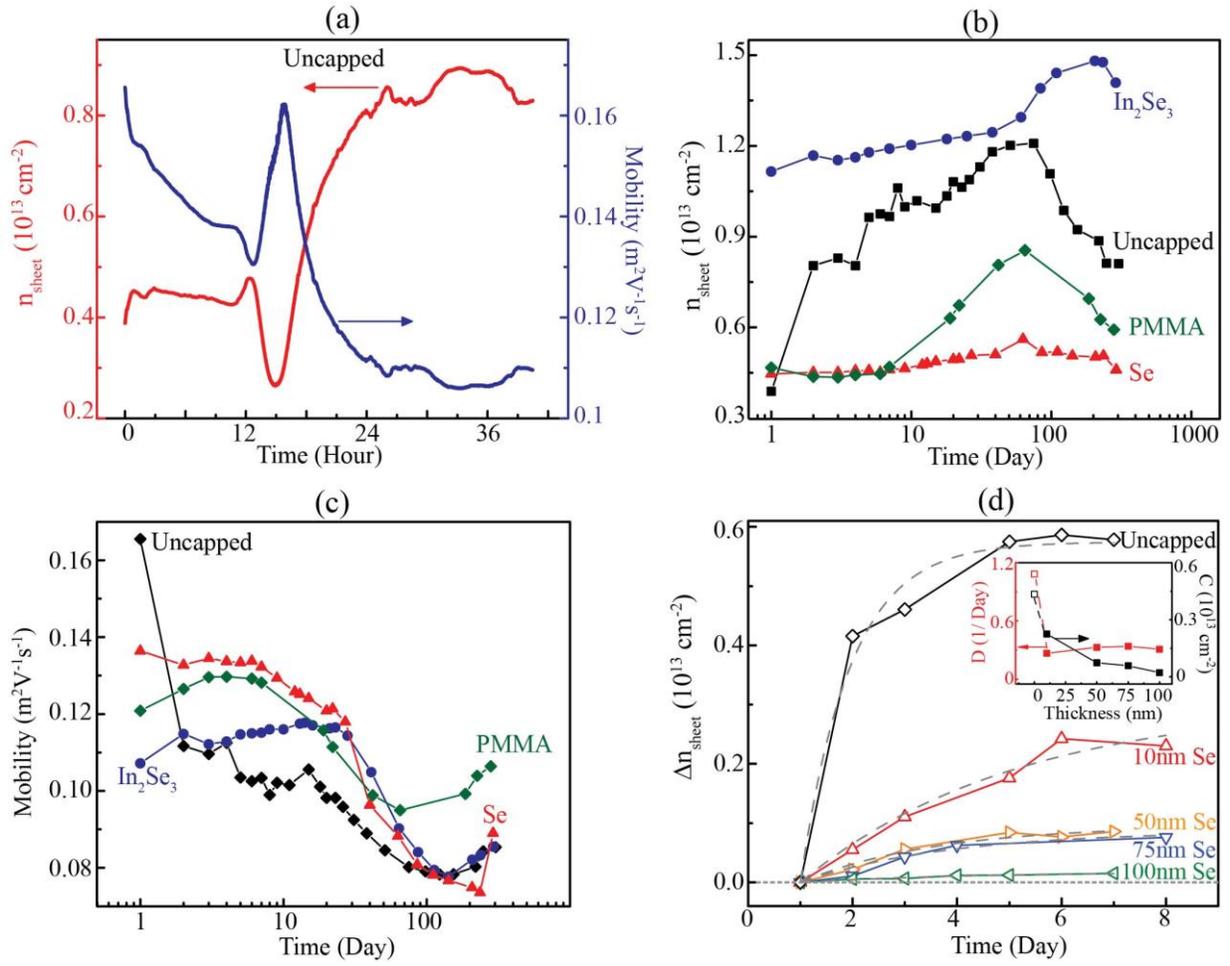

**FIG. 2 (2column)**

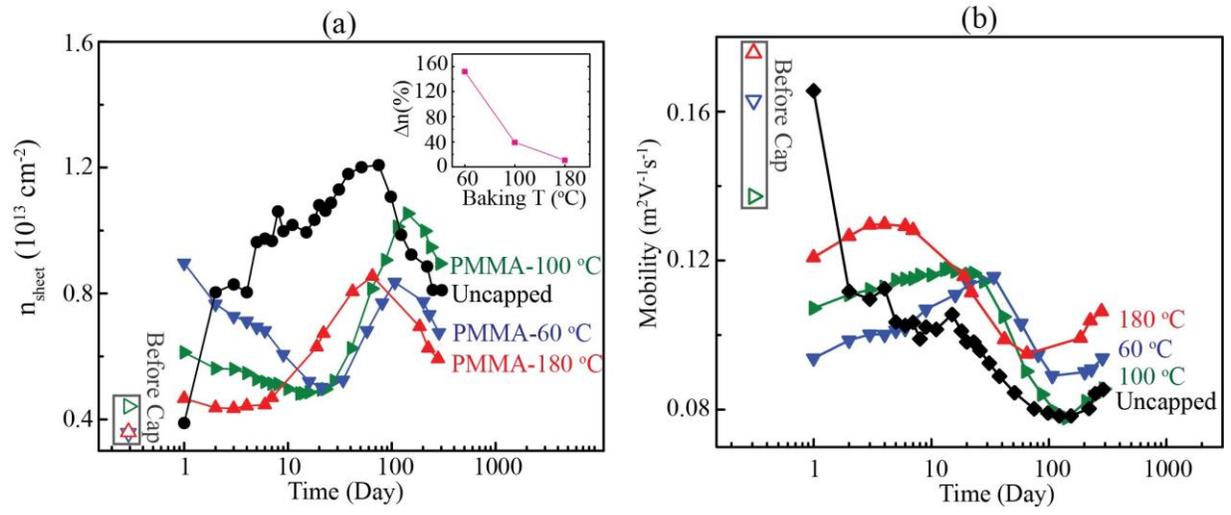

**FIG. 3 (2column)**

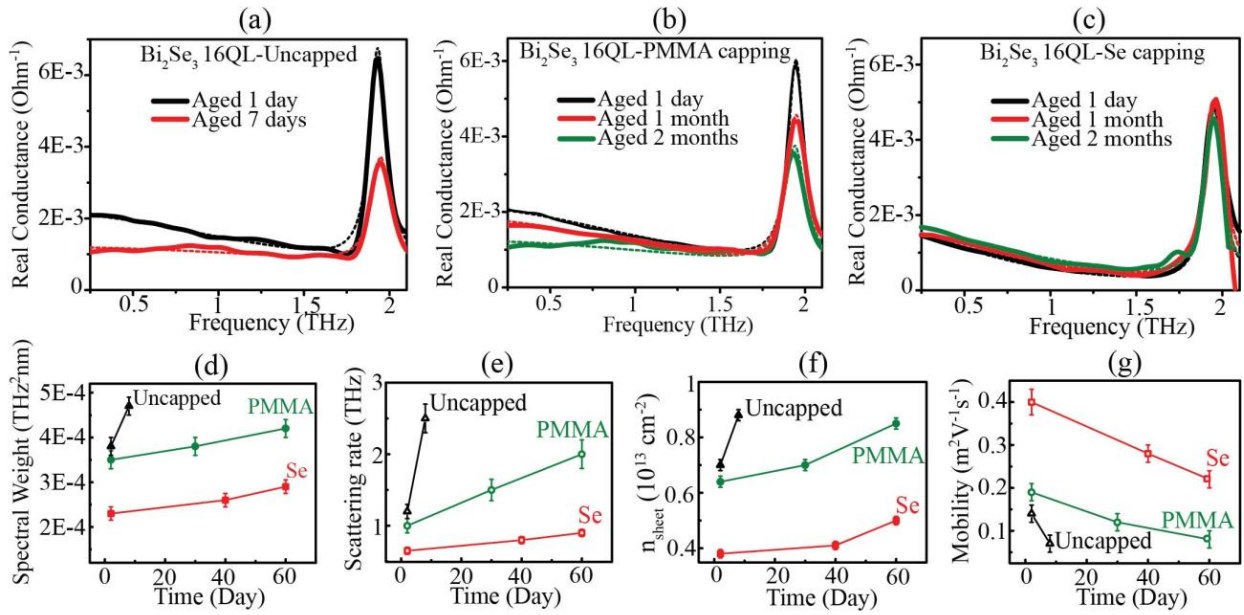